\documentclass[twoside,english,1p]{elsarticle}
\usepackage[T1]{fontenc}
\usepackage[latin9]{inputenc}
\usepackage{float}
\usepackage{amsbsy}
\usepackage{amssymb}
\usepackage{graphicx}
\usepackage{esint}

\makeatletter
\journal{Comptes Rendus de Physique}

\usepackage{filemod}

\newcommand{\kbar}{\mathchar'26\mkern-9mu k}

\newcommand{\noshow}[1]{}

\renewcommand\[{\begin{equation}}
\renewcommand\]{\end{equation}}
\renewenvironment{eqnarray*}{\begin{eqnarray}}{\end{eqnarray}}
\usepackage[ddmmyyyy]{datetime}

\makeatother

\usepackage{babel}
\begin{document}

\begin{frontmatter}{}

\title{Quantum simulation of disordered systems with cold atoms}

\author{Jean Claude Garreau}

\ead{jean-claude.garreau@univ-lille1.fr}

\address{Universit\'e de Lille, CNRS, UMR 8523 \textendash{} PhLAM \textendash{}
Laboratoire de Physique des Lasers Atomes et Mol\'ecules, F-59000
Lille, France}
\begin{abstract}
This paper reviews the physics of quantum disorder in relation with
a series of experiments using laser-cooled atoms exposed to ``kicks''
of a standing wave, realizing a paradigmatic model of quantum chaos,
the kicked rotor. This dynamical system can be mapped onto a tight-binding
Hamiltonian with pseudo-disorder, formally equivalent to the Anderson
model of quantum disorder, with quantum chaos playing the role of
disorder. This provides a very good \emph{quantum simulator} for the
Anderson physics.
\end{abstract}

\end{frontmatter}{}
\begin{center}
{\footnotesize{}(File: \jobname) Dated: \filemodprint{\jobname}}\\
{\small{}(v. 2.1, compilation: \today)}
\par\end{center}{\small \par}

\section{Introduction: When paradigms meet\label{sec:Introduction}}

Disorder and chaos are ubiquitous phenomena in the macroscopic world,
and are often intertwined. In the quantum world, however, they are
completely distinct, because (classical) chaotic dynamics relies on
nonlinearities, while the quantum world is governed by the \emph{linear}
Schr\"odinger (or Dirac) equation. A widely accepted definition of
\emph{quantum chaos} is: The behavior of a system \emph{whose classical
counterpart is chaotic}. Because of the linearity of Schr\"odinger
equation, quantum systems cannot show \emph{sensitivity to the initial
conditions}, \emph{sine qua non} condition of classical chaos. More
quantitatively, quantum dynamics never produces positive Lyapunov
exponents: Quantum ``chaotic'' dynamics is thus \emph{qualitatively
different} from classical chaos. However, ``signatures'' of the
classical chaotic behavior might show up on the behavior of a quantum
system, they are the ``imprint'' of the existence of a positive
Lyapunov exponent in the corresponding classical system. One of the
best known of these signatures is level repulsion, implying that the
energy level-spacing distribution tends to zero when the spacing tends
to zero~\citep{Haake:QuantumSigChaos:01}.

Quantum disorder has been intensively investigated for almost 60 years
since P. W. Anderson introduced his paradigmatic model~\citep{Anderson:LocAnderson:PR58},
describing (in a somewhat crude, but mathematically tractable, way)
the quantum physics of disordered media. The model's main prediction
is the existence of exponentially-localized eigenstates in space,
in sharp contrast with the delocalized Bloch eigenstates of a prefect
crystal. In three dimensions (3D), the model predicts the existence
of a second-order quantum phase transition between delocalized (``metal'')
and localized (``insulator'') phases, known as the \emph{Anderson
metal-insulator transition}, which is the main subject of the present
work.

The \emph{kicked rotor} is a paradigm of Hamiltonian classical and
quantum chaos. The classical version of this simple system displays
a wealth of dynamic behaviors; which makes it well adapted for studies
of quantum chaos. Surprisingly, in its quantum version, the classical
chaotic diffusion in momentum space can be totally suppressed, leading
to an exponential localization in \emph{momentum} space, called \emph{dynamical
localization}~\citep{Casati:LocDynFirst:LNP79}, which strongly evokes
Anderson localization. Indeed, it has been proved~\citep{Fishman:LocDynAnderson:PRA84}
that there exists a mathematical mapping of the kicked rotor Hamiltonian
onto an 1D Anderson Hamiltonian. The kicked rotor can thus be used
to \emph{quantum-simulate} Anderson physics.

The original idea of \emph{quantum simulation} seems to be due to
Feynman~\citep{Feynman:SimulatingPhysics:IJTP82}. Basically, the
idea of a quantum simulator (in today's sense, which is somewhat different
from Feynman's) is to realize a physical model originally introduced
in a certain domain using a ``simulator'' from another domain, which
presents practical or theoretical advantages compared to the original
model. The present work describes a quantum simulator obtained by
the encounter of two paradigms: The Anderson model, originally proposed
in a condensed-matter context, is simulated using the atomic kicked
rotor formed by laser-cooled atoms interacting with laser light. Another
beautiful example of quantum simulation is the realization of Bose-
(or Fermi-) Hubbard physics with ultracold atoms, e.g. the observation
of the Mott transition~\citep{Greiner:MottTransition:N02}; other
examples can be found in refs.~\citep{Bloch:ManyBodyUltracold:RMP08,Georgescu:QuantumSimulation:RMP14}. 

\section{The Anderson model in a nutshell\label{sec:Anderson-model}}

\begin{figure}
\begin{centering}
\includegraphics[width=6.3cm]{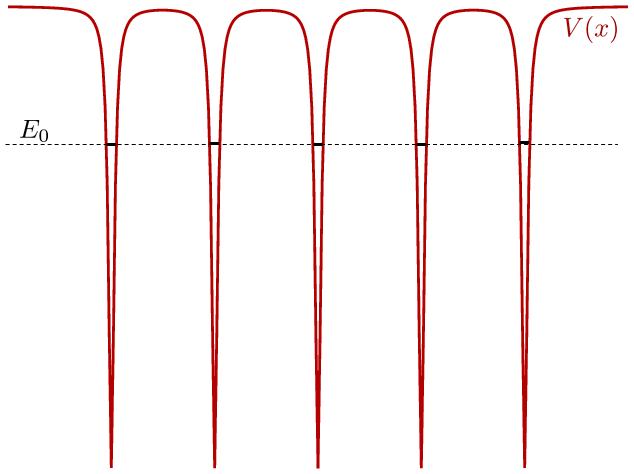}$\quad$\includegraphics[width=6.3cm]{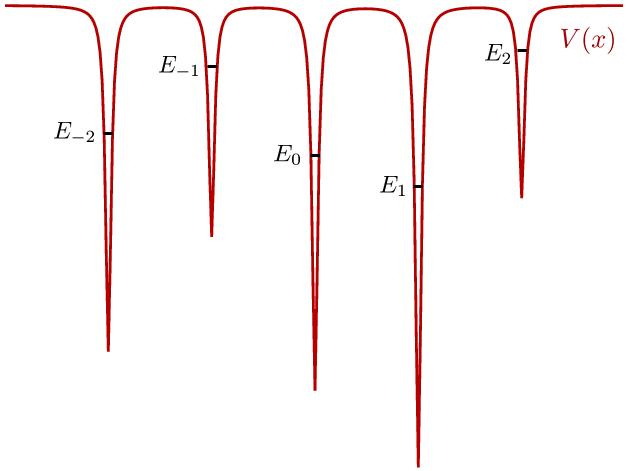}
\par\end{centering}
\caption{\label{fig:TBAnderson}Left: Tight-binding model of a perfect crystal.
Energy levels associated to individual sites have identical energies.
Right: Disordered ``crystal'', the site energies have a random distribution.}

\end{figure}

One can have a taste of the Anderson model by considering a 1D crystal
in a tight-binding description~\citep{Dalibard:ReseauxOptiques3:LiaisonsFortes:13}.
Figure~\ref{fig:TBAnderson} (left) shows schematically the tight-binding
description of a crystal: The electron wave function is written in
a basis of states localized in individual potential wells, usually
the so-called Wannier states $w_{n}(x)\equiv\left\langle x\right.\left|n\right\rangle $~\citep{Wannier:WannierStates:PR37,Dalibard:ReseauxOptiques3:LiaisonsFortes:13}
where $\left|n\right\rangle $ is the basis ket corresponding to site
$n$. Supposing that the temperature is low enough that only the ground
state of each well is in play here, the translation symmetry implies
\emph{i}) that the levels corresponding to each well all have the
same energy $E_{0}$ and \emph{ii}) that $w_{n}(x)=w_{0}(x-n)$. Tunneling
between wells a distance $r$ apart (in units of the lattice constant)
in a Hamiltonian of the form $H=p^{2}/2m+V(x)$ (with $V(x+n)=V(x)$,
$n\in\mathbb{Z}$) is thus allowed, with an amplitude $T(n,n+r)=\int w_{n}^{*}(x)Hw_{n+r}(x)dx$$=\int w_{0}^{*}(x)Hw_{r}(x)dx\equiv T_{r}$,
so that the Hamiltonian can be written
\[
H_{\mathrm{TB}}=\sum_{n\in\mathbb{Z}}\left(E_{0}\left|n\right\rangle \left\langle n\right|+\sum_{r\in\mathbb{Z}^{*}}T_{r}\left|n\right\rangle \left\langle n+r\right|\right).
\]
The first (``diagonal'') term describes the on-site energy and the
second (``hopping'') one describes tunneling between sites a distance
$r$ apart. The equation for an eigenstate $u_{\epsilon}(x)=\sum_{n}u_{n}w_{n}(x)$
in position representation is thus
\begin{equation}
E_{n}u_{n}+\sum_{r\neq0}T_{r}u_{n+r}=\epsilon u_{n},\label{eq:HAnderson}
\end{equation}
with, in the present case, $E_{n}=E_{0}$.

Anderson \emph{postulated}~\citep{Anderson:LocAnderson:PR58} that
the main effect of disorder in such systems is to randomize the on-site
energy {[}Fig.~\ref{fig:TBAnderson} (right){]}, the effect on the
coupling coefficients $T_{r}$ being ``second-order''; this hypothesis
is called ``diagonal disorder''. Anderson used on-site energies
given by a simple box distribution $E_{n}\in[-W/2,W/2]$. One can
heuristically understand the existence of localization in such lattice
by the following argument. Tunneling is reduced because in presence
of diagonal disorder the energy levels are not degenerate, but it
is not completely suppressed: An electron can still ``hop'' to a
neighbor site presenting an energy defect $\Delta E$, but can stay
there only for a time $\tau\sim\hbar/\Delta E$, after which it should
come back to the initial site or jump to a more ``favorable site''
with a smaller energy defect. The order of magnitude of $\tau$ is
$\hbar/\Delta E$$\sim\hbar/W$, and the average time for a hop to
a neighbor site is $\hbar/T$, where $T$ is the typical value of
$T_{r}$, one thus expects that if $T\ll W$ (strong disorder) the
electron essentially stays close to its initial site: This is the
physical origin of the Anderson localization. In this case, it turns
out that the eigenstates are exponentially localized $u_{\epsilon}(x)=\sum_{n}u_{n}(\epsilon)w_{n}(x)$$\sim\exp\left(-\left|x-x_{0}\right|/\xi\right)$,
where $\xi(\epsilon)$ is the \emph{localization length}.

On the other hand, if $T\gg W$ (weak disorder), one can expect the
electron to diffuse in the crystal. How far can it go? The non-trivial
answer, obtained using renormalization group arguments by Anderson
and coworkers~\citep{Abrahams:Scaling:PRL79}, twenty years after
the original article, is that it depends on the dimension. In 1D,
it can be shown that the eigenstates are \emph{always localized},
whatever the value of the disorder $W$ and of the eigenenergy $\epsilon$~\citep{Luck:SystDesord:92,MuellerDelande:DisorderAndInterference:arXiv10}.
In 3D all eigenstates are localized for strong enough disorder ($W/T\ge16.5$),
but for smaller disorder there is co-existence of delocalized states
(if $\left|\epsilon\right|<E_{c}(W)$, where $E_{c}$ is the so-called
the \emph{mobility edge}) and localized states~\footnote{In~\citep{Anderson:LocAnderson:PR58} Anderson hints at the possible
co-existence of localized and delocalized states by this prudent phrase:
``We can show that a typical perturbed state is localized with unit
probability; but we cannot prove that it is possible to assign localized
perturbed states a one-to-one correspondence with localized unperturbed
states in any obvious way, so that perhaps with very small probability
a few states may not be localized in any clear sense''.} (see Fig.~1 of ref.~\citep{Kroha:SelfConsistentTheoryAnderson:PRB90}).
This 3D transition between localized (insulator) and diffusive (metal)
states is called the \emph{Anderson metal-insulator transition}, which
is the main subject of our quantum simulations\emph{.} 

One can also give a heuristic explanation for the existence of this
transition using scaling arguments~\citep{Abrahams:Scaling:PRL79}.
Consider small cubes of a material, each one presenting a ``resistance''
$r$ to transport (as such cubes are taken as a unit of length, $r$
plays the role of a resistivity, depending on the microscopic properties
of the material). First consider a stack where the cubes form a 1D
line. In the linear (ohmic) regime, if the number of cubes is $L$,
the total resistance is $R\sim rL$: The larger the stack, the larger
the resistance. Hence, the resistance is always finite and there is
no diffusion in 1D in the limit $L\to\infty$. Now consider the case
where the small cubes are arranged in 3D, large cube of size $L$
thus containing $L^{3}$ small cubes: Then, the total resistance in
the direction of the flow is proportional to $rL$ as before, but
\emph{must be divided by the transverse area}, that is $R\sim rL/L^{2}\sim L^{-1}$,
which tends to zero as $L\to\infty$, so that in large enough \emph{3D}
stacks the transport is \emph{diffusive}. Generalizing such argument,
the resistance of the macroscopic cube of size $L$ in dimension $d$
is $R\sim rL/L^{d-1}$$=rL^{2-d}$. For large values of $r$, the
linear (ohmic) approximation breaks down, and one has $R\sim r^{L}/L^{2}$,
which tends to infinity as $L\to\infty$, and the transport is suppressed.
There are thus two asymptotic regimes: \emph{i}) A large $r$ regime
where $R$ always increases with $L$ and thus becomes an insulator
for a large enough stack, and \emph{ii}) A low $r$ regime where $R$
scales as $L^{2-d}$, and may diverge (if $d<2$) or vanish (if $d>2$)
when $L\to\infty$. Hence, for $d>2$ there must be (supposing that
$R(L)$ is a smooth function) some value of the resistivity $r=r_{c}$
separating a region of diffusive transport (for $r<r_{c}$) of a region
where transport is suppressed (for $r>r_{c}$). Up to now, we have
made no hypothesis on the \emph{microscopic} origin of the resistivity
$r$. In the particular case where the resistivity is due to disorder,
the suppression of transport is associated to the Anderson localization.
The 2D case presents a marginal behavior with $R\sim L^{0}$ in the
ohmic regime, thus independent of $L$; 2D is a ``pathological''
dimension (known as the \emph{lower critical dimension}) for which
there is always localization but with a localization length increasing
exponentially as $r\to0$.

The Anderson transition is a second order quantum phase transition,
characterized by two critical exponents: $\xi\sim\left(W_{c}-W\right)^{\nu}$
on the insulator side $W<W_{c}$, and $D\sim\left(W-W_{c}\right)^{s}$
($D$ is the diffusion coefficient) in the metal $W>W_{c}$ side,
but it turns out that in 3D $\nu=s$ (``Wegner's law''~\citep{Wegner:ScalingMobilityEdge:ZFP76}).
Numerical simulations of the 3D Anderson model show that $\nu\sim1.57$~\citep{Slevin:AndersonCriticalExp:NJP2014}
which puts the Anderson transition in the ``orthogonal'' universality
class of time-reversal-invariant systems, to which it indeed belongs,
as the Hamiltonian~(\ref{eq:HAnderson}) is time independent. 

Experimental study of the Anderson model in condensed matter is difficult,
for a variety of reasons: The tight-binding model is a one-electron
approach, which does not take into account electron-electron interactions;
the ions are supposed static, which means very low temperatures; decoherence
is supposed negligible, etc. Moreover, it is difficult to obtain direct
information on the electron wave function in a crystal, and thus to
directly observe Anderson localization. This is typically a situation
where quantum simulation can be useful. Can one study Anderson physics
with other systems? In fact this has been done even before the notion
of quantum simulation became popular, by noting that this physics
can be observed (with more or less difficulty) with any kind of waves
propagating in disordered media, as light~\citep{Maret:AndersonTransLight:PRL06,Wiersma:LightLoc:N97,Schwartz:LocAnderson2DLight:N07}
or acoustic waves~\citep{Faez:MultifractAnderson:PRL09}. However,
these wave systems have their own difficulties, the main one being
that fluorescence and absorption tend to have the same exponential
signature as localization~\citep{Sperling:Can3dLightLocalizationBeReached:NJP16}.

In this respect, ultracold atom systems are excellent quantum simulators.
One can generate a disordered ``optical potential'' \textendash{}
which the atoms see as a mechanical potential (see sec.~\ref{sec:KickedRotor})
\textendash{} by passing laser light through a depolished plate, producing
a ``speckle''. With this kind of system, Anderson localization has
been observed in 1D~\citep{Billy:AndersonBEC1D:N08} (or Anderson-like~\citep{Roati:AubryAndreBEC1D:N08})
and 3D~\citep{Jendrzejewski:AndersonLoc3D:NP12,Semeghini:MobilityEdgeAnderson:NP15},
and anisotropic localization has also been observed in a 3D geometry
with a degenerate Fermi gas~\citep{Kondov:ThreeDimensionalAnderson:S11}~\footnote{In practice this is not so easy, as 3D Anderson localization is observed
if the so-called Ioffe-Regel criterion is fulfilled, that is if $k\ell\sim1$
where $\ell$ is the mean free path of the atoms in the optical potential.
One thus needs a speckle with the shortest possible correlation length.
Moreover, as $k$ is the de Broglie wave number of the atoms, one
needs ultracold atoms. Also, it is difficult to produce an \emph{isotropic}
disordered optical potential.}. These are beautiful experiments, but the determination of the mobility
edge is challenging, because of the dependence of the localization
length on the energy (which is hard to control inside the speckle).
The critical exponent of the transition could not yet been measured
with such systems.

The quantum simulator described above, obtained by placing ultracold
atoms in a speckle potential, is a ``direct translation'' of the
Anderson model in cold-atom language. Are there other models that
are mathematically equivalent to the Anderson model without being
a direct translation? Indeed yes, the atomic kicked rotor meets these
conditions, and presents considerable interest for the simulation
of the Anderson transition, as we shall see in the next sections.

\section{The kicked rotor and the Anderson model\label{sec:KickedRotor}}

\subsection{The kicked rotor and the dynamical localization}

In its simplest version, the kicked rotor (KR) is formed by a particle
constrained to move on a circular orbit to which periodic (period
$T_{1}$) delta-pulses (\emph{kicks}) of a constant force are applied.
As only the component of the force along the orbit influences the
particle's motion, one has the Hamiltonian
\begin{equation}
H_{\mathrm{kr}}=\frac{L^{2}}{2}+K\cos\theta\sum_{n\in\mathbb{Z}}\delta(t-n)\label{eq:Hkr}
\end{equation}
where $(L,\theta)$ are conjugate variables corresponding to the angular
momentum and the angular position of the particle, and we use units
in which the momentum of inertia is 1, time units of the period $T_{1}$,
and $K\cos\theta$ is the torque applied at each kick. The corresponding
Hamilton equations can be easily integrated over one period, giving
a stroboscopic map for each $t=n^{+}$, known as the \emph{Standard
Map}:
\[
\theta_{n+1}=\theta_{n}+L_{n},\qquad L_{n+1}=L_{n}+K\sin\theta_{n+1}.
\]
This very simple map \textendash{} easy to implement numerically \textendash{}
displays a wealth of dynamical behaviors going from regular ($K\ll1$),
to mixed (chaos and regular orbits, $1\lesssim K\lesssim5$) and to
developed (ergodic) chaos ($K>5$). In the last case, the motion is
simply a diffusion in momentum space: $\overline{L^{2}}=2Dt$ (the
overbar denotes a classical average in phase space). This sequence
of behaviors for increasing $K$ very closely follows the Kolmogorov-Arnol'd-Moser
(KAM) scenario~\citep{Chirikov:ChaosClassKR:PhysRep79}. This has
made the Standard Map a paradigm for studies of classical Hamiltonian
chaos.

It is straightforward to write the Schr\"odinger equation for the
KR Hamiltonian~(\ref{eq:Hkr}). However, for periodic systems, the
one-period evolution operator is often more useful:
\begin{equation}
U(1)=\exp\left(-i\frac{L^{2}}{2\hbar}\right)\exp\left(-i\frac{K\cos\theta}{\hbar}\right)=\exp\left(-i\frac{m^{2}\hbar}{2}\right)\exp\left(-i\frac{K\cos\theta}{\hbar}\right),\label{eq:Ukr}
\end{equation}
where, in the second expression, we used the quantization of angular
momentum $L=m\hbar$. Although $\theta$ and $L$ do not commute,
the presence of the delta function allows one to neglect $L^{2}/2\hbar$
compared to $K\cos\theta\delta(0)/\hbar$ in the argument of the second
exponential, so that $U$ effectively factorizes into a ``free evolution''
and a ``kick'' part. One can efficiently simulate such an evolution,
As the kick part is diagonal in the position space and the free propagation
part in the momentum space, the evolution over a period ``costs''
only two Fourier transforms and two multiplications. This makes the
kicked rotor a privileged ground for studies of quantum chaos. The
typical action integrated over one period is, in units of $\hbar$,
$L^{2}T_{1}/\hbar\sim m^{2}T_{1}$, so that by adjusting $T_{1}$
one controls the ``quantum character'' of the dynamics. More precisely,
by increasing $T_{1}$ one reduces the time scale for quantum effects
to become dominant (the so-called \emph{Heisenberg time}); one can
thus think of an ``effective Planck constant'' $\hbar_{\mathrm{eff}}\propto T_{1}$.
In the momentum representation, the kick operator is proportional
to $\sum_{j}J_{j}(K/\hbar_{\mathrm{eff}})$$\exp\left(ij\theta\right)$$\left|m+j\right\rangle \left\langle m\right|$
($J_{j}(x)$ is the Bessel function of first type and order $j$),
its application thus generates ``side bands'' in the momentum distribution
within a range $K/\hbar_{\mathrm{eff}}$. After a few kicks, the amplitude
for a given angular momentum $m\hbar$ is the sum of such contributions,
generating an interference that leads to purely quantum effects.

The first numerical study of the \emph{quantum} kicked rotor (QKR)~\citep{Casati:LocDynFirst:LNP79}
produced a surprising result: In the ergodic regime $K>5$ a classical
diffusion in momentum space was observed for short times, but for
later times the kinetic energy $\left\langle L^{2}\right\rangle /2$
was observed to saturate at a constant value~\footnote{In~\citep{Casati:LocDynFirst:LNP79} the authors write ``We do not
yet understand this quantum anomaly''.}. At the same time, the momentum distribution was observed to change
from a Gaussian to an exponential $\exp\left(-\left|m\right|/\xi\right)$,
which hints to a relation to the Anderson localization. This phenomenon
was called ``dynamical localization'', i.e. localization in the
momentum space.

\subsection{The atomic kicked rotor\label{subsec:AtomicKR}}

Graham \emph{et al}.~\citep{Graham:LDynTheo:PRA92} first suggested
using cold atoms for observing dynamical localization, and the first
experimental observation was made by Raizen and coworkers in 1994~\citep{Moore:LDynFirst:PRL94},
with a somewhat different system. In later experiments~\citep{Moore:AtomOpticsRealizationQKR:PRL95},
this same group used an atomic realization of the QKR, obtained by
placing laser-cooled atoms in a far-detuned standing wave. In such
conditions, the atoms see the radiation as a sinusoidal mechanical
potential \textendash{} called an optical or dipole potential \textendash{}
affecting their center of mass motion. If the radiation is periodically
pulsed ($T_{1}\sim30$~$\mu$s~\footnote{We indicate typical values used in our setup for the parameters.})
with very short pulses~\footnote{The typical time scale of the atom dynamics should be much larger
than the pulse duration, in practice this means pulses of a few hundred
ns. }, then the corresponding Hamiltonian is 
\begin{equation}
H_{\mathrm{akr}}=\frac{p^{2}}{2\mu}+K\cos x\sum_{n}\delta(t-n)\label{eq:Hakr}
\end{equation}
which has the same form as~(\ref{eq:Hkr}). Distances are measured
in units of $\lambda_{L}/4\pi$, where $\lambda_{L}=2\pi/k_{L}$ is
the laser wavelength, time in units of the pulse period $T_{1}$,
and $K\propto T_{1}I/\Delta$, where $I\sim10$~W/cm$^{2}$ is the
radiation intensity and $\Delta\sim\pm20$~GHz$\approx3\times10^{3}\Gamma$
($\Gamma$ is the natural width of the transition) its detuning with
respect to the atomic transition. The reduced Planck constant is in
this case $h_{\mathrm{eff}}\equiv$$\kbar=4\hbar k_{L}^{2}T_{1}/M$$\sim2.9$,
where $M$ is the mass of the atom. By convention, we shall use sans
serif characters to indicate \emph{dimensioned} quantities, e.g. $x=2k_{L}\mathsf{x}$,
$t=\mathsf{t}/T_{1}$, etc. It is useful to chose units such that
$\mu=\kbar^{-2}$ in~(\ref{eq:Hakr}), so that $p=\mathsf{p}/2\hbar k_{L}$.
The lattice constant is $\lambda_{L}/2$, comparable to the de Broglie
wavelength of laser-cooled atoms $\sim\lambda_{L}/3$, ensuring the
quantum character of the dynamics. The laser-atom detuning must be
large enough that the typical spontaneous emission time $\sim\left(\Gamma I/\Delta^{2}\right)^{-1}$
is larger than the duration of the experiment, otherwise the random
character of the phase changes induced by the spontaneous emission
process produces lethal decoherence effects~\citep{Nowak:DelocalizationfUltracoldLightScatt:PRA12,Cohen:LocDynTheo:PRA91}.
Because the momentum exchanges between the standing wave and the atom
are directed along $\boldsymbol{k}_{L}$, the dynamics is effectively
1D, the transverse directions, not being affected by the radiation,
evolve independently of the longitudinal direction.

\begin{figure}[h]
\begin{centering}
\includegraphics[width=8cm]{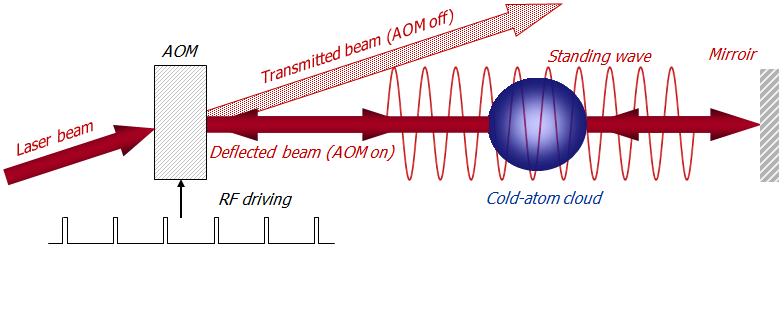}
\par\end{centering}
\begin{centering}
\includegraphics[width=10cm]{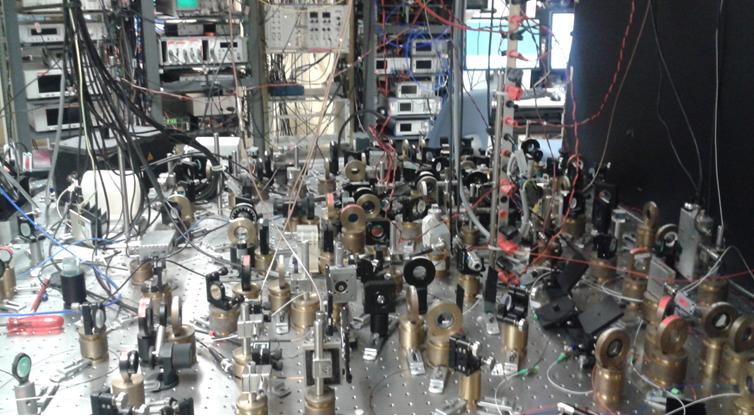}
\par\end{centering}
\caption{\label{fig:ExperimentalSetup}Experimental setup. Top: Schematic view:
An acousto-optical modulator (AOM) is driven by a pulsed radio-frequency
(RF) signal. If the RF is on (pulses) the AOM deflects the laser beam
which is reflected backwards by a mirror and generates a standing
wave that interacts with the atomic cloud. If the RF driving is off,
the undeflected beam does not interact with the atoms. Bottom: ``Less
schematic'' view of the experiment.}
\end{figure}

Our basic experimental setup is shown schematically in Fig.~\ref{fig:ExperimentalSetup}
(top): Cesium atoms are laser-cooled in a magneto-optical trap to
temperatures $\sim2\mu$K. The (dissipative) trap is then turned off,
and pulses of a far-detuned standing wave applied to the atoms (in
such conditions, the resulting dynamics is quantum). The final momentum
distribution of the atoms can be measured by a variety of techniques.
In a first version of the experiment, the standing wave was horizontal,
to avoid gravitation acceleration, but this limited the interaction
time of the freely-falling atoms with the standing wave. In more recent
times, we used a vertical standing wave formed by two frequency-chirped
laser beams to compensate gravity: By shifting the frequency of one
beam with respect to the other, we obtain a standing wave whose nodes
are accelerated in the vertical direction. By adjusting this acceleration
so that it equals the gravity acceleration, a kicked rotor is obtained
in the \emph{free-falling} reference frame~\citep{Manai:Anderson2DKR:PRL15}.
Our group used for many years stimulated Raman transitions between
hyperfine sublevels~\citep{Ringot:RamanSpectro:PRA01,Chabe:Polarization:OC07},
as we changed to the vertical standing wave configuration, we could
use a standard time-of-flight technique. 

\begin{figure}
\begin{centering}
\includegraphics[width=6.3cm]{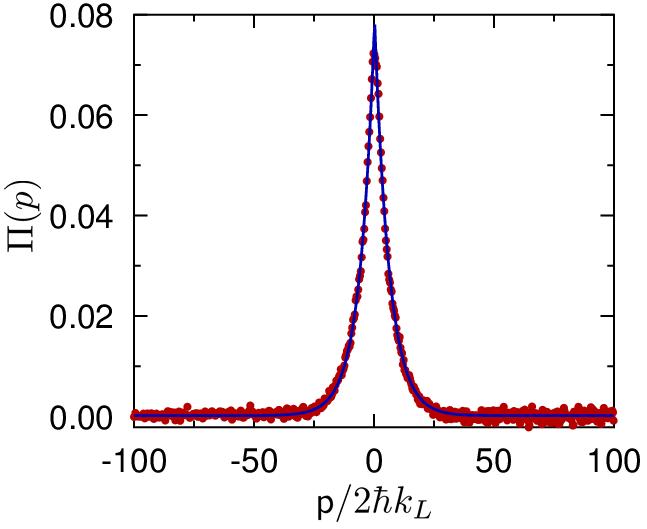}$\qquad$\includegraphics[width=6.3cm]{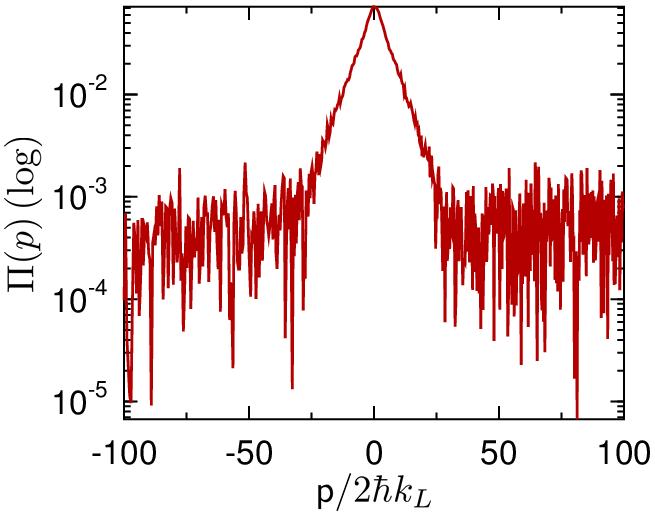}
\par\end{centering}
\caption{\label{fig:ExpDL}Experimental observation of dynamical localization:
Exponentially-localized momentum distribution $\Pi(p)$ recorded at
$t=30$ kicks. The left plot in linear scale shows a fit by an exponential
shape (blue curve). The right plot in semilog scale confirms the exponential
localization. Parameters are $K=7.2$, $\kbar=2.89$.}

\end{figure}

Because of the spatial periodicity of the optical potential, quasimomentum
is a constant of motion. Hence kicks couple only momentum components
separated by $2\hbar k_{L}$, so that starting with a well defined
momentum $\mathsf{p}_{0}$ leads to a discrete momentum distribution
at values $(m+\beta)2\hbar k_{L}$ with $m\in\mathbb{Z}$ and $\beta$
the fractional part of $p_{0}$. The main difference between this
``unfolded'' kicked rotor and the ``standard'' version described
by the Hamiltonian~(\ref{eq:Hkr}) is the existence of quasimomentum
families. One can show (see sec.~\ref{subsec:FGP}) that each quasimomentum
family maps to a particular Anderson eigenvector with a given disorder~\footnote{Taking quasimomentum into account, Eq.~(\ref{eq:MappingEn}) becomes
$E_{n}=\tan\left(\omega/2-(m+\beta)^{2}\kbar/4\right)$.}. This is quite useful in many situations, as averaging over quasimomentum
(that is, starting with an initial state that has a width comparable
to the Brillouin zone width $2\hbar k_{L}$) is equivalent to average
over disorder in the corresponding Anderson model. In other situations,
however, this averaging can hide interesting effects. Typically, the
momentum distribution of \emph{cold} atoms (that is, of atoms cooled
in a magneto-optical trap) populate a few Brillouin zones, but that
of \emph{ultracold} atoms (e.g. a Bose-Einstein condensate) can be
only a small fraction of the Brillouin zone. In general, interatomic
interactions are negligible, if not from the start, after a few kicks,
as the spatial density dilutes very quickly due to the diffusion in
the momentum space. An experimentally observed momentum distribution
is shown in Fig.~\ref{fig:ExpDL}. 

\subsection{Equivalence between dynamical localization and the 1D Anderson localization\label{subsec:FGP}}

The observation that the momentum distribution takes an exponential
shape in dynamical localization strongly evokes Anderson localization.
A few years after the discovery of dynamical localization, Fishman,
Grempel and Prange established a mathematical equivalence between
the two~\citep{Fishman:LocDynAnders:PRL82,Fishman:LocDynAnderson:PRA84}.
This equivalence is basically a matter of algebra, but as it is the
ground for our using of the KR as a quantum simulator of Anderson
physics, it is useful to see how it arises.

As the KR is periodic in time, the Floquet operator technique can
be used, which consists in diagonalizing the one-period evolution
operator~(\ref{eq:Ukr}): 
\begin{equation}
\exp\left(-i\frac{p^{2}}{2\kbar}\right)\exp\left(-i\frac{K\cos x}{\kbar}\right)\left|\omega\right\rangle =e^{-i\omega}\left|\omega\right\rangle \label{eq:Floquet}
\end{equation}
where $\left|\omega\right\rangle $ is the Floquet ``quasi-eigenstate'',
and $\omega\in[0,2\pi)$ is the corresponding ``quasi-energy''.
The ``stroboscopic'' evolution of any initial state $\left|\psi_{0}\right\rangle $
at integer times $t=n$ is then simply: 
\[
\left|\psi_{n}\right\rangle =\sum_{\omega}e^{-i\omega n}\left\langle \omega\right.\left|\psi_{0}\right\rangle \left|\omega\right\rangle .
\]
Using this expansion, one can write
\[
\left\langle p^{2}\right\rangle =\sum_{\omega}\left|\left\langle \omega\right.\left|\psi_{0}\right\rangle \right|^{2}\left\langle \omega\right|p^{2}\left|\omega\right\rangle +\sum_{\omega\neq\omega^{\prime}}\left\langle \psi_{0}\right.\left|\omega^{\prime}\right\rangle \left\langle \omega\right.\left|\psi_{0}\right\rangle e^{i\left(\omega^{\prime}-\omega\right)n}\left\langle \omega^{\prime}\right|p^{2}\left|\omega\right\rangle .
\]
If the Floquet spectrum is dense (which happens if the system is in
the quantum-chaotic regime) and $n=\mathsf{t}/T_{1}$ is large enough
that $\left(\omega^{\prime}-\omega\right)n\ge\pi/2$ the contributions
in second term tend to interfere destructively and only the first
sum $\omega^{\prime}=\omega$ survives: 
\begin{equation}
\left\langle p^{2}\right\rangle \to p_{\mathrm{\infty}}^{2}=\sum_{\omega}\left|\left\langle \omega\right.\left|\psi_{0}\right\rangle \right|^{2}\left\langle \omega\right|p^{2}\left|\omega\right\rangle \quad(t>t_{\mathrm{l\mathrm{oc}}})\label{eq:Asymptoticp2}
\end{equation}
after some localization time $t_{\mathrm{loc}}$. The kinetic energy
thus tends to saturate for $t>t_{\mathrm{loc}}$, as observed numerically
by Casati \emph{et al}.~\citep{Casati:LocDynFirst:LNP79}.

As dynamical localization manifests itself in momentum space, it is
natural to express the Floquet states in the momentum basis. However,
the kick operator is not diagonal in this representation, so Fishman,
Grempel, and Prange used the trigonometric identity
\begin{equation}
e^{\mathrm{i}x}=\frac{1+i\tan(x/2)}{1-i\tan(x/2)}\label{eq:Identity}
\end{equation}
to transform equation~(\ref{eq:Floquet}) into
\begin{equation}
\frac{1+i\tan\hat{v}}{1-i\tan\hat{v}}\left(1-i\tan\hat{t}\right)\frac{1}{1+i\tan\hat{t}}\left|\omega\right\rangle =\left|\omega\right\rangle \label{eq:FGP}
\end{equation}
with $\hat{v}=\omega/2-\hat{p}^{2}/4\hbar$ and $\hat{t}=\kappa\cos\hat{x}/2$
($\kappa:=K/\kbar$). Decomposing $\left(1+i\tan\hat{t}\right)^{-1}\left|\omega\right\rangle =\sum_{s}u_{s}\left|s\right\rangle $
on the basis momentum eigenstate $\left|s\right\rangle $ gives $\left|\omega\right\rangle =$$\left(1+i\tan\hat{t}\right)\sum_{s}u_{s}\left|s\right\rangle $.
Projecting~(\ref{eq:FGP}) on a momentum eigenstate $\left\langle m\right|$
and using the previous identity leads, after some (cumbersome) algebra,
to
\begin{equation}
\tan\left(v_{m}\right)u_{m}+\sum_{r\neq0}t_{r}u_{m+r}=-t_{0}u_{m},\label{eq:FGPeqeigenvalues}
\end{equation}
where $v_{m}=\omega/2-m^{2}\kbar/4$ and $t_{r}=-\left\langle m\right|\tan\hat{t}\left|m+r\right\rangle $,
which has the form of the 1D tight-binding equation~(\ref{eq:HAnderson})
with the equivalences
\begin{equation}
E_{n}\leftrightarrow\tan v_{m}=\tan\left(\frac{\omega}{2}-\frac{m^{2}\kbar}{4}\right)\label{eq:MappingEn}
\end{equation}
\begin{equation}
T_{r}\leftrightarrow t_{r}=\frac{1}{2\pi}\int_{0}^{2\pi}dxe^{irx}\tan\left(\kappa\cos x/2\right).\label{eq:MappingTr}
\end{equation}
Note that the kick term in the evolution operator basically maps on
the hopping coefficient $t_{r}\sim K/\kbar$ that controls the transport,
while the free propagation term maps on the equivalent of the diagonal
disorder, $\tan v_{m}$, which is essentially controlled by $\kbar$.
Hence the Anderson control parameter $T/W$ translates into $K/\kbar$
to within a numerical factor.

There are however differences with respect to the Anderson model: 

1) In~(\ref{eq:FGPeqeigenvalues}), all eigenstates $u_{m}$ correspond
to the \emph{same} Anderson eigenvalue $\epsilon=-t_{0}$ (according
to~(\ref{eq:MappingTr}), by symmetry, $t_{0}=0$). This in particular
means that all localized states for the KR have the \emph{same} localization
length $\xi$, in contrast to the 1D Anderson eigenstates, whose localization
length scales as $(T/W)^{2}$~\citep{Luck:SystDesord:92,MuellerDelande:DisorderAndInterference:arXiv10}.
This fact has an important consequence: If, in~(\ref{eq:Asymptoticp2}),
the width of the initial state $\psi_{0}(p)$ (supposed centered at
$p=0$) is much smaller than $\xi$, this state will be projected
only over Floquet eigenstates spreading over a range $\sim\xi$; one
thus concludes that $p_{\infty}^{2}\approx p_{\mathrm{loc}}^{2}\sim\xi^{2}=\mathrm{cte},$
i.e. the asymptotic value of the kinetic energy does not depend on
the initial state~\footnote{Moreover, the QKR does not map onto a nearest-neighbor Anderson model.
The nearest-neighbor approximation is usual in the Anderson model
context, but it is not necessary: It suffices that the hopping coefficients
$t_{r}$ decreases fast enough (at least as $r^{-3}$) to ensure the
convergence of the perturbation series used by Anderson (see~\citep{Anderson:LocAnderson:PR58}).}. 

\begin{figure}[H]
\begin{centering}
\includegraphics[height=6cm]{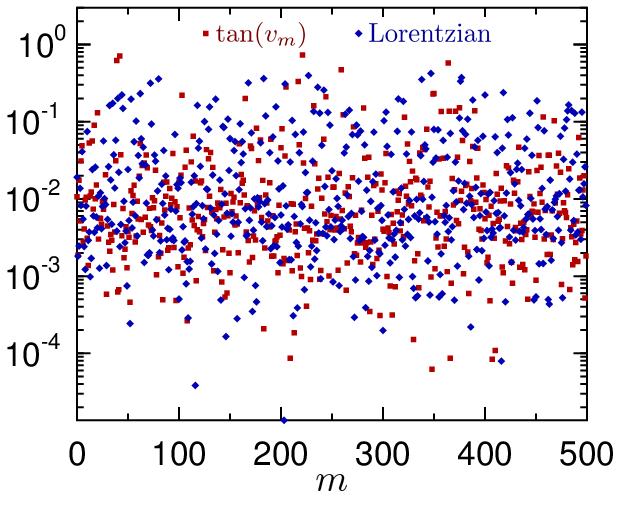}
\par\end{centering}
\caption{\label{fig:PseudoRandom}Pseudo-random distribution (log scale) of
``on-site energies'' for the KR compared to a Lorentzian distribution.}
\end{figure}

2) The Anderson model random on-site energies map into the \emph{deterministic}
function $\tan v_{m}$. However, for large enough values of $\kbar\propto T_{1}$,
this functions varies strongly (Fig.~\ref{fig:PseudoRandom}), at
the condition that $\kbar$ is an \emph{irrational number}, otherwise
the function is periodic in $m$~\footnote{If $\kbar$ is rational, a totally different \textendash{} but also
interesting \textendash{} behavior arises, called ``quantum resonance'',
namely a \emph{ballistic} increase of the kinetic energy $\left\langle p^{2}\right\rangle \propto t^{2}$
(for rational values of the quasimomentum $\beta$).}.

3) The localized states $\left|u\right\rangle =\sum_{m}u_{m}\left|m\right\rangle =\left(1+i\tan\hat{t}\right)^{-1}\left|\omega\right\rangle $
are \emph{not} the Floquet states, indeed $\left|\omega\right\rangle =2e^{\mathrm{i}t}$$\left(e^{i\hat{t}}+e^{-\mathrm{i}\hat{t}}\right)$$\left|u\right\rangle $.
As $\hat{t}$ is ``half'' the kick operator, the Floquet state is,
to within a phase factor, a superposition of the localized state advanced
and retard of half a kick; it has thus essentially the same localization
properties.

The 1D Anderson localization length for fixed energy is proportional
to $(T/W)^{2}$~\citep{Luck:SystDesord:92,MuellerDelande:DisorderAndInterference:arXiv10}
and the fact that the dynamics is diffusive for $t<t_{\mathrm{loc}}$
with an early-time (classical) diffusion constant $D\approx K^{2}/4$,
so that $p_{\mathrm{loc}}^{2}\sim2Dt_{\mathrm{loc}}$, imply that
both $t_{\mathrm{loc}}$ and $p_{\mathrm{loc}}$ are proportional
to $(K/\kbar)^{2}$.

\subsection{The quasiperiodic kicked rotor and its equivalence to a 3D Anderson
model~\label{subsec:Equivalence3D}}

In order to study the Anderson transition, one needs an analog of
the \emph{3D} Anderson model. A simple idea would be to use a 3D kicked
rotor, by kicking a 3D optical lattice~\citep{Wang:AndersonKR3D:PRE09}.
This is experimentally complicate for several reasons, the main ones
being that gravity breaks the symmetry among directions and the delicate
control of the phase relations of the various beams forming a separable
3D optical lattice.

In refs.~\citep{Shepelyansky:Bicolor:PD87,Casati:IncommFreqsQKR:PRL89},
it was suggested that a $d$-dimensional tight-binding Hamiltonian
with pseudo-disorder can be obtained by adding $d-1$ new incommensurate
frequencies to a spatially 1D kicked rotor. Specifically, let us consider
the Hamiltonian
\begin{equation}
H_{\mathrm{qpkr}}=\frac{p^{2}}{2\mu}+K\cos x\left(1+\varepsilon f(t)\right)\sum_{n}\delta(t-n)\label{eq:Hqpkr}
\end{equation}
where $f(t)$ is a modulation function presenting frequencies $\omega_{2}...\omega_{d}$.
If all $\omega_{i}/2\pi$ are rational numbers, the driving is periodic,
and the system localizes, with a different localization time. If some
of the frequencies $\omega_{i}/2\pi$ are irrational, the Fishman-Grempel-Prange~\citep{Fishman:LocDynAnderson:PRA84}
procedure described in sec.~\ref{subsec:FGP} cannot be applied,
because, not being periodic in time, Hamiltonian~(\ref{eq:Hqpkr})
does not admit Floquet states. For simplicity, we shall consider here
the $d=2$ case, that is $f(t)=\cos\left(\omega_{2}t+\varphi_{2}\right)$
with $\omega_{2}/2\pi\in\mathbb{R}\backslash\mathbb{Q}$; the generalization
to higher dimensions is immediate. We introduce an ``intermediary''
2D \emph{periodic} Hamiltonian in an extended Hilbert space $\mathcal{S}\otimes\mathcal{S}_{2}$
where $\mathcal{S}$ is the ``real'' Hilbert space corresponding
to the $(x,p)\equiv(x_{1},p_{1})$ degree of freedom (where the atoms
live) and $\mathcal{S}_{2}$ is a ``virtual'' space corresponding
to formal variables $(\hat{x}_{2},\hat{p}_{2})$, defining a ``virtual''
degree of freedom:
\begin{equation}
H_{\mathrm{kr2D}}=\frac{p^{2}}{2\mu}+\omega_{2}\hat{p}_{2}+K\cos x\left(1+\varepsilon\cos\hat{x}_{2}\right)\sum_{n}\delta(t-n)\label{eq:Hkr2D}
\end{equation}
(we temporarily put a $\hat{}$ on operators in the virtual dimension
to make the argument clearer). In the Hilbert space $\mathcal{S}\otimes\mathcal{S}_{2}$
one can define a unitary transformation $T(t)=\exp\left(i\omega_{2}t\hat{p}_{2}\right)$
(corresponding to a rotating frame with frequency $\omega_{2}$ around
the direction 1) that transforms the above Hamiltonian into
\begin{eqnarray}
H_{\mathrm{kr2D}}^{\prime} & = & TH_{\mathrm{kr2D}}T^{\dagger}+i\frac{dT}{dt}T^{\dagger}\nonumber \\
 & = & \frac{p^{2}}{2\mu}+K\cos x\left[1+\varepsilon\cos\left(\hat{x}_{2}+\omega_{2}t\right)\right]\sum_{n}\delta(t-n).\label{eq:Hkr2Dprime}
\end{eqnarray}
If we now consider a \emph{restriction} in $\mathcal{S}_{2}$ to the
states generated by $T(t)\left|\varphi_{2}\right\rangle $ (for all
$t$), with $\left|\varphi_{2}\right\rangle $ being a position eigenstate
in $\mathcal{S}_{2}$, $\hat{x}_{2}\left|\varphi_{2}\right\rangle =\varphi_{2}\left|\varphi_{2}\right\rangle $,
we obtain Eq.~(\ref{eq:Hqpkr}) {[}with $f(t)=\cos\left(\omega_{2}t+\varphi_{2}\right)${]}.
This means that the evolution generated by~(\ref{eq:Hqpkr}) and~(\ref{eq:Hkr2D})
are \emph{identical} provided that the initial state in $\mathcal{S}_{2}$
is \emph{any fixed eigenstate of the position}~\footnote{Note that because of the linear dependence of $H_{\mathrm{kr2D}}$
on $\hat{p}_{2}$ the evolution is dispersionless, so that a system
prepared in a well-defined position $\left\langle x_{2}\right.\left|\varphi_{2}\right\rangle =\delta(x_{2}-\varphi_{2})$
stays perfectly localized. }. For the Hamiltonian~(\ref{eq:Hqpkr}) this is equivalent to say
that the kick modulation phase $\varphi_{2}$ is well defined, which
is the case experimentally up to the very good precision of the synthesizer
generating this frequency. To obtain a $d$-dimension Hamiltonian,
one simply takes e.g. $f(t)=\prod_{i=2}^{d}\cos\left(\omega_{i}t+\varphi_{i}\right)$,
starts from a generalized Hilbert space $\mathcal{S}\otimes\mathcal{S}_{2}\otimes...\otimes\mathcal{S}_{d}$,
and defines the restriction accordingly.

Hamiltonian~(\ref{eq:Hkr2D}) \emph{is periodic in time} (of period
$T_{1}$), and thus \emph{has} Floquet quasi-eigenstates to which
the Fishman-Grempel-Prange mapping \emph{can} be applied. The algebra
is completely analogous to that of sec.~\ref{subsec:FGP} and results
in a $d$-dimensional Anderson eigenvalue equation
\begin{equation}
\tan v_{\boldsymbol{m}}u_{\boldsymbol{m}}+\sum_{\boldsymbol{r}\neq0}t_{\boldsymbol{r}}u_{\boldsymbol{m}+\boldsymbol{r}}=-t_{\boldsymbol{0}}u_{\boldsymbol{m}},\label{eq:HAnderson-1}
\end{equation}
where $\boldsymbol{m}$ and $\boldsymbol{r}$ are now vectors in $\mathbb{Z}^{d}$.
Equations~(\ref{eq:MappingEn}) and~(\ref{eq:MappingTr}) generalize
to
\begin{eqnarray}
\tan v_{\boldsymbol{m}} & = & \tan\left(\omega/2-m^{2}\kbar/4-(m_{2}\omega_{2}+...+m_{d}\omega_{d})/2\right)\label{eq:pseudodisorder-d}\\
t_{\boldsymbol{r}} & = & \frac{1}{\left(2\pi\right)^{d}}\int_{0}^{2\pi}d\boldsymbol{x}e^{i\boldsymbol{r}\cdot\boldsymbol{x}}\tan\left[\left(\kappa/2\right)\cos x\left(1+\varepsilon\cos x_{2}...\cos x_{d}\right)\right],\label{eq:hopping-d}
\end{eqnarray}
which correspond to a $d$-dimensional pseudo-disorder, provided that
$\kbar$, $\omega_{2}$,...,$\omega_{d}$ and $2\pi$ are co-prime
numbers \footnote{The fact that the modulation frequencies must be co-prime with 2$\pi$
is, as in the $d=1$ case (sec.~\ref{subsec:FGP}), the condition
for the quasiperiodicity of the pseudo-disorder in all directions
$m_{2},...,m_{d}$. If two of the frequencies have a rational relation,
say $\omega_{2}/\omega_{3}=p/q$, one can set $\bar{m}=qm_{2}+pm_{3}$,
the pseudo-disorder is characterized by a \emph{single} integer $\bar{m}$
instead of $m_{2},m_{3}$ and its effective dimension is thus reduced
by one. An interesting question is what should be the condition between
the frequencies and $\kbar$. A tentative argument is as follows:
In the $d=2$ case for simplicity, suppose that $\omega_{2}/\kbar=p/q$
with $p$ and $q$ co-prime integers. Then one can write $m^{2}\kbar+2m_{2}\omega_{2}$$=\kbar\left(qm^{2}+2m_{2}p\right)/q$
and define $\bar{m}=qm^{2}+2m_{2}p$. Obviously, not all integers
are of the form $\bar{m}$, but one can define a pseudo-disorder $\tan v_{\ell}$
given by Eq.~(\ref{eq:pseudodisorder-d}) if $\ell$ is of the form
$\bar{m}$, and equal to some fixed value $\bar{\epsilon}$ otherwise.
This pseudo-disorder is again characterized by a single integer, and
is thus ``effectively'' one dimensional. Note this is only a \emph{definition}
of a \emph{particular} pseudo-disorder of dimension 1, which does
not apply to the hopping coefficients, that are still given by Eq.~(\ref{eq:hopping-d}):
We are \emph{not} trying to map the 2D lattice onto a 1D lattice,
but to construct a 2D lattice with 1D pseudo-disorder.}. This equivalence allows one to quantum-simulate the Anderson model
in any dimension!

\begin{figure}
\begin{centering}
\includegraphics[height=6cm]{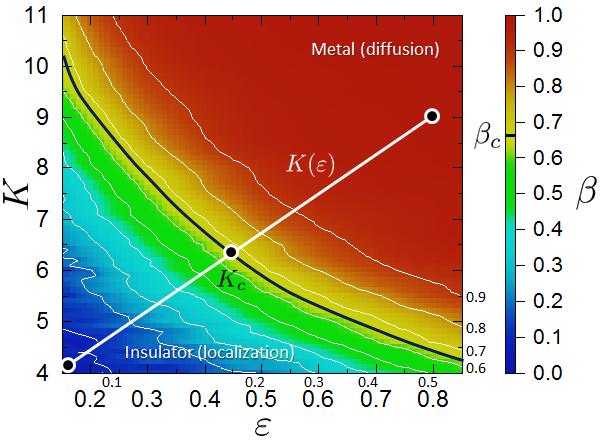}
\par\end{centering}
\caption{\label{fig:PhaseDiag}Phase diagram of the quasiperiodic KR Anderson
transition in the $(\varepsilon,K)$ plane. The quantity $\beta=d\ln\left\langle p^{2}\right\rangle /d\ln t$
at $t=10^{4}$, plotted in false colors, is a direct measure of the
transport properties. The critical line corresponds to $\beta_{c}=2/3$,
and is indicated by the black line in the diagram. The white line
illustrates a path crossing the transition often used in our experiments.
(Numerical data by D. Delande).}
\end{figure}

For $d=3$ the above model allows the observation of the Anderson
transition (see sec.~\ref{subsec:AndersonTransition}). Figure~\ref{fig:PhaseDiag}
displays the phase diagram of such a transition in the $(\varepsilon,K)$
plane, calculated numerically. This has been first evidenced numerically
in~\citep{Shepelyansky:Bicolor:PD87,Casati:IncommFreqsQKR:PRL89}
and experimentally with a quasiperiodic atomic KR by our group~\citep{Chabe:Anderson:PRL08}.
As in the periodic case, all KR states map to the same Anderson eigenvalue
$-t_{\boldsymbol{0}}$. This has an interesting consequence: There
is no mobility edge in this transition, all Floquet quasi-states are
either localized or diffusive according to the choice of the parameters
$(\varepsilon,K)$. This simplifies considerably the determination
of the critical exponent of the transition as compared to ``directly
mapped'' ultracold-atom quantum simulators~\citep{Jendrzejewski:AndersonLoc3D:NP12,Semeghini:MobilityEdgeAnderson:NP15}.

\subsection{Theory of the Anderson localization}

There is no complete theory of the Anderson localization. In his original
article~\citep{Anderson:LocAnderson:PR58}, using the Hamiltonian~(\ref{eq:HAnderson})
in 3D, Anderson basically sums a Green function perturbation series
and evaluates the probability of finding the electron at a distance
of $n$ sites from a given initial site, which turns out to be exponentially
small if $W/T\gtrsim16.5$~\footnote{Anderson's paper overestimates this threshold}. 

Anderson's work inspired various attempts to sum the perturbation
series (or analogous series) using diagrammatic methods~\citep{AkkermansMontambaux:MesoscopicPhysics:11,Rammer:QuantumTransport:04}.
Taking into account only the simplest loops to evaluate the ``return
to the origin'' probability, this approach shows a \emph{reduction}
of the quantum diffusion coefficient with respect to the classical
one, an effect known as \emph{weak localization}\footnote{Weak localization has measurable physical effects, the best known
being a factor 2 enhancement of the probability for a wave (sound,
light, matter waves) in a disordered media to be scattered in the
direction opposite to its propagation, called \emph{coherent backscattering}
\emph{effect}.}\emph{,} in contrast with \emph{strong} \textendash{} or Anderson
\textendash{} localization, for which the quantum diffusion coefficient
is strictly zero. In the diagrammatic theory of weak localization,
the modified diffusion coefficient $D_{q}$ is expressed as~\footnote{Complications as the fact that $D$ is in fact a tensor are ignored
here for simplicity.}
\begin{equation}
D_{q}=D_{cl}\left(1-\frac{C}{\rho}\int\frac{d\boldsymbol{q}}{D_{cl}\boldsymbol{q}^{2}}\right)\label{eq:WL}
\end{equation}
where $D_{cl}$ is the classical diffusion coefficient, $\boldsymbol{q}=\hbar\boldsymbol{k}$
is the momentum and $\rho$ the density of states and $C$ is a constant.
The formula is valid only if the contribution of higher-order loops
is negligible, that is if $\left|D_{cl}/D_{q}-1\right|\ll1$, thus
it cannot describe strong localization.

A possible improved approximation consists in considering that $D_{cl}$
in the denominator of the integrand should itself be corrected, and
a direct way to do so is to replace it by $D_{q}$. Eq.~(\ref{eq:WL})
then becomes an implicit equation that has to be solved self-consistently;
this approach is thus called the \emph{self-consistent theory} of
the Anderson localization. The theory presents several difficulties,
one of them being that the integral is not convergent, and one has
to set appropriate cut-offs based on physical considerations. 

Despite that, these theories allow qualitative understanding and quantitative
predictions, as e.g. the existence of a phase transition in 3D and
the calculation of the localization length $\xi$. The main drawback
is that the self-consistent theory predicts a critical exponent $\nu=1$
for the transition, not agreeing with the numerical value $\nu\approx1.57$~\citep{Slevin:AndersonCriticalExp:NJP2014}
(for time-reversal-invariant systems). These methods are thus useful
tools whose results should however be used with care. D. Delande,
G. Lemari\'e and N. Cherroret have obtained several interesting results
by transposing these theories to the quasiperiodic kicked rotor~\citep{Lemarie:UnivAnderson:EPL09,Lemarie:These:09,Cherroret:AndersonNonlinearInteractions:PRL14}.

More complex theories are based on field-theoretical approaches, like
supersymmetry, first developed by Efetov~\citep{Efetov:SupersymmetryInDisorder:97}
and used in many relevant works~\citep{Altland:FieldTheoryQKR:PRL96,Tian:TheoryAndersonTransition:PRL11,Chen:QuantumDrivenIntegerQuantumHallEffect:PRL14,Tian:EmergencerQuantumHallEffectChaos:PRB16}.

\section{Quantum simulation of disordered systems with the kicked rotor}

The first quantum simulation of (1D) Anderson localization with matter
waves is due to Raizen and co-workers. In 1994 they observed localization
in a driven cold-atom system~\citep{Moore:LDynFirst:PRL94} and,
a little latter, with an atomic kicked rotor~\citep{Moore:AtomOpticsRealizationQKR:PRL95}.
In 1998 our group started the development of the experiment described
in sec.~\ref{subsec:AtomicKR} for studies of the quantum chaos.
Quite soon, we realized that more complex temporal driving was the
key ingredient for richer dynamics, as illustrated in our early papers~\citep{Ringot:Bicolor:PRL00,Szrift:SubFourier:PRL02,Lignier:Reversibility:PRL05,Chabe:PetitPic:PRL06}.
A long term effort of experimental improvements and better theoretical
understanding allowed us to start quantum simulations of the Anderson
transition. These efforts resulted in a rather complete study of the
Anderson transition, which is still under investigation. In this section,
I present the more prominent features of these studies; the interested
reader can find more details in the corresponding publications.

\subsection{Anderson transition, phase diagram and critical exponent\label{subsec:AndersonTransition}}

\begin{figure}
\begin{centering}
\includegraphics[height=6cm]{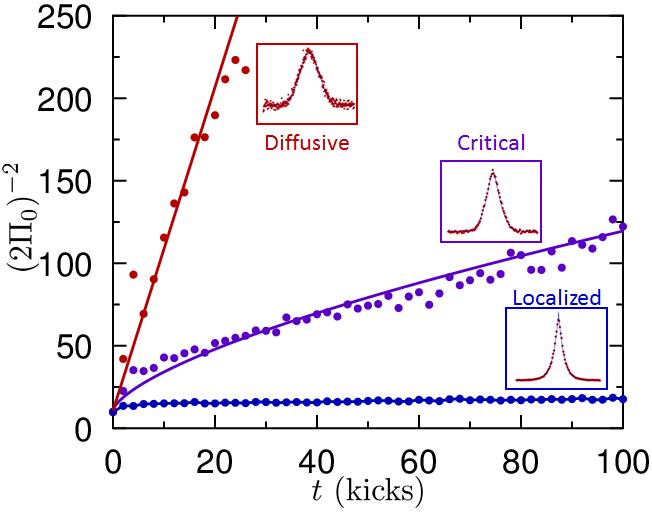}
\par\end{centering}
\caption{\label{fig:AndersonTrans}Experimental observation of the Anderson
transition. The blue curve (dots are experimental points, solid lines
are fits) indicates kinetic energy $\left(2\Pi_{0}\right)^{-2}\propto\left\langle p^{2}\right\rangle $
in the localized regime, the violet one in the critical regime $p^{2}\propto t^{2/3}$,
and the red one in the diffusive regime $p^{2}\propto t$. The insets
show momentum distributions at 150 kicks, with the corresponding fits,
$\Pi(p)\propto\exp\left(-\left|p\right|^{\alpha}/s\right)$ with $\alpha=$1,
3/2, 2 respectively. Parameters are, resp.: $K=4,\varepsilon=0.35$,
$K=6.3,\varepsilon=0.55$, $K=9,\varepsilon=0.8$. For all curves
$\omega_{2}/2\pi=\sqrt{5}$, $\omega_{3}/2\pi=\sqrt{13}$ and $\kbar=2.89$.}

\end{figure}

The Anderson transition manifests itself in 3 or more dimensions (cf.~sec.~\ref{sec:Anderson-model}).
Its quantum simulation with the kicked rotor relies on a Hamiltonian
of the form discussed in sec.~\ref{subsec:Equivalence3D}, with two
incommensurate additional frequencies:
\[
H_{\mathrm{kr3D}}=\frac{p^{2}}{2\mu}+K\cos x\left[1+\varepsilon\cos(\omega_{2}t+\varphi_{2})\cos(\omega_{3}t+\varphi_{3})\right]\sum_{n}\delta(t-n).
\]
From such a Hamiltonian, a numerical phase diagram can be constructed,
as shown in Fig.~\ref{fig:PhaseDiag}. Observing the dynamics in
different parts of the diagram is, in principle, not difficult. One
can rely on the shape of the momentum distribution: Exponential (i.e.
localized) in the lower-left part of the diagram or Gaussian (i.e.
diffusive) in the upper-right part. One can also use the average second-momentum
$\left\langle p^{2}\right\rangle $, which displays characteristic
asymptotic behaviors $p_{\mathrm{loc}}^{2}$ (localized) and $\propto t$
(diffusive). Experimentally, it is easier to measure $\Pi_{0}(t)$,
the zero-momentum class population, which is, to within a factor of
order of one, $\left(2\left\langle p^{2}\right\rangle \right)^{-1/2}$.
This is shown in Fig.~\ref{fig:AndersonTrans}.

A scaling argument allows one to understand the properties of the
critical behavior. In the parameter plane shown in Fig.~\ref{fig:PhaseDiag},
a path (white line) crossing the critical curve (black line) is described
by some function $K(\varepsilon)$, and to each value of $K$ (or
$\varepsilon$) a curve $\left\langle p^{2}\right\rangle (t)$ can
be associated. Defining $x\equiv K-K_{c}$, the scaling hypothesis
implies that $\left\langle p^{2}\right\rangle (x,t)=t^{m}f(xt^{\mu})$
for any $(x,t)$, where $f$ is a universal scaling function (i.e.
independent of the microscopic details as $\kbar,\omega_{2},\omega_{3}$,
etc.). Close to and below the transition point $(x\rightarrow0^{-})$,
dynamical localization implies that $\left\langle p^{2}\right\rangle (x,t\rightarrow\infty)\rightarrow p_{\mathrm{loc}}^{2}$.
The existence of a second order phase transition, on the other hand,
implies that $p_{\mathrm{loc}}\sim x^{-\nu}$ for $x\rightarrow0^{-}$,
where $\nu$ is the critical exponent on the transition's insulator
side. As $p_{\mathrm{loc}}^{2}=t^{m}(xt^{\mu})^{-2\nu}$ is independent
of $t$, one should have $m-2\mu\nu=0$. The same reasoning applies
on the metallic side $x\rightarrow0^{+}$, where $\left\langle p^{2}\right\rangle (x,t)\rightarrow2Dt$
and $D\sim x^{s}$, where $s$ is the critical exponent on the transition's
metallic side, then $\left\langle p^{2}\right\rangle (x,t)=t^{m}(xt^{\mu})^{s}\propto t$
implies $m+\mu s=1$. Finally, the so-called Wegner's law~\citep{Wegner:ScalingMobilityEdge:ZFP76}
states that $s=(d-2)\mu$ where $d>2$ is the dimension (thus $s=\nu$
for $d=3$). This univocally determines $m=2/d$. At the critical
point $x=0$,
\begin{equation}
\left\langle p^{2}\right\rangle (0,t)=t^{2/d}f(0)\label{eq:criticalp2}
\end{equation}
 and, for $d=3$, one retrieves the critical behavior $\left\langle p^{2}\right\rangle \propto t^{2/3}$
observed in Fig.~\ref{fig:AndersonTrans}. 

\begin{figure*}
\begin{centering}
\includegraphics[width=4.2cm]{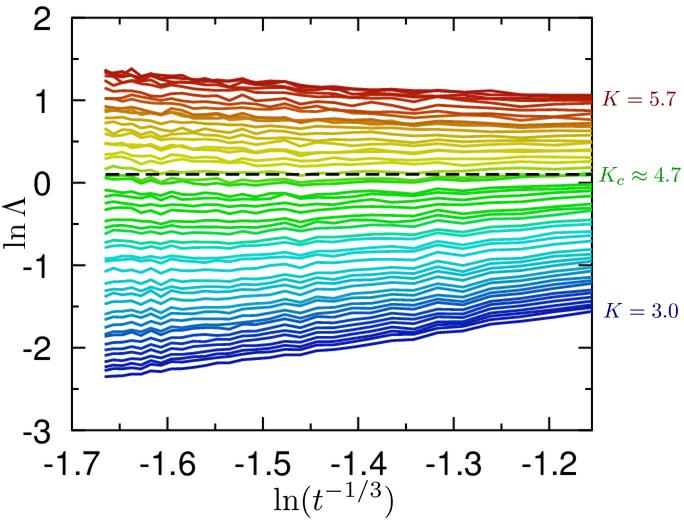}$\quad$\includegraphics[width=4.2cm]{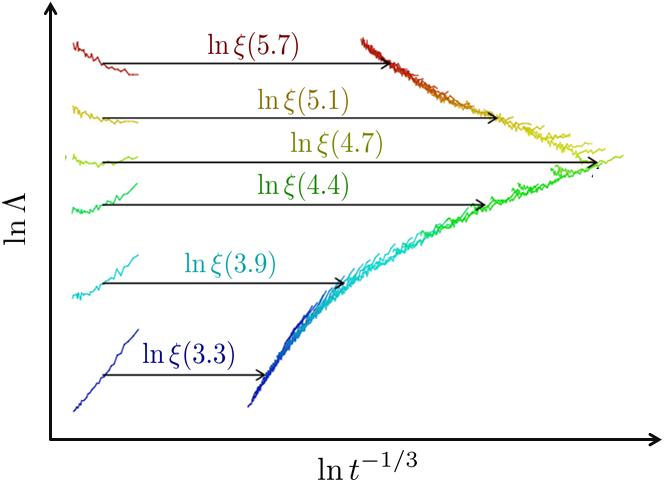}$\quad$\includegraphics[width=4.2cm]{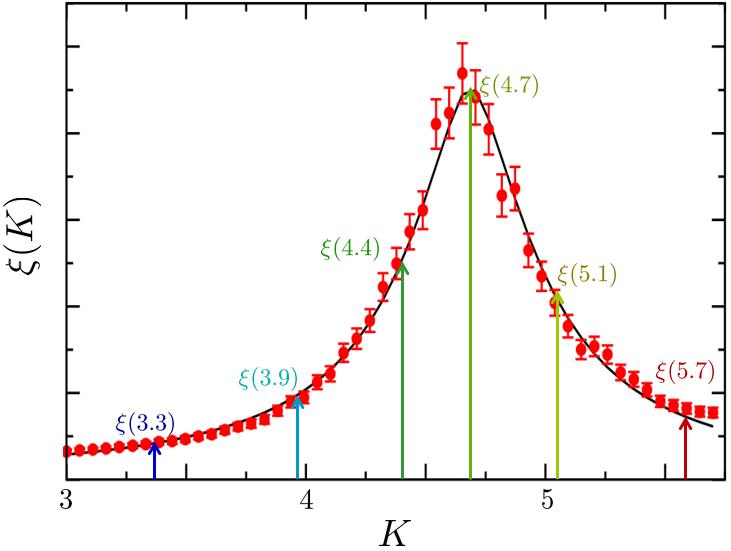}
\par\end{centering}
\caption{\label{fig:Scaling}Finite-\emph{time} scaling. \emph{Left:} The quantity
$\Lambda(t^{-1/3})\equiv\left(t^{-1/3}\right)^{2}\left\langle p^{2}\right\rangle (K-K_{c},t)$
is plotted for various values of $3\le K\le5.7$. By construction,
the horizontal curve (dotted line) corresponds to the transition point
$K=K_{c}\approx4.7$ for the particular path $K(\varepsilon)$ chosen
here. \emph{Center}: A few curves are used to illustrate the principle
of the finite-\emph{time} scaling method: Each curve can be translated
by a quantity $\ln\xi(K)$, and the aim is to obtain a curve $f(\xi t^{-1/3})$
continuous everywhere except at the critical point. One thus obtains
a lower ``localized'' branch of slope 2 (blue-cyan-green) and an
upper ``diffusive'' branch (yellow-orange-red) of slope -1. The
``filaments'' one sees detaching from the curves are a signature
of the presence of decoherence in the system. \emph{Right}: Collecting
the values of $\xi(K)$ one constructs the scaling factor, which allows
the determination of the critical exponent $\nu$ through a fit with
an algebraic function with a cutoff (cf.~text).}
\end{figure*}

The relevant scaling quantity is thus $\Lambda(x)\equiv t^{-2/3}\left\langle p^{2}\right\rangle (x,t)$,
whose characteristic behavior close to the transition is: $\Lambda(0^{-})\sim x^{-2\nu}t^{-2/3}$$=x^{-2\nu}\left(t^{-1/3}\right)^{2}$,
$\Lambda(0)=\mathrm{cte}$ and $\Lambda(0^{+})=x^{\nu}t^{1/3}$$=x^{\nu}\left(t^{-1/3}\right)^{-1}$.
Fig.~\ref{fig:Scaling} (left) shows a set of $\Lambda$ functions
obtained from our experimental data for different values of $(\varepsilon,K)$
across the transition. The scaling hypothesis then implies $\Lambda(x,t^{-1/3})=f\left(\xi(x)t^{-1/3}\right)$
where $f$ is a continuous (except at $x=0$) scaling function and
$\xi(x)$ is a scaling factor to be determined. In the log-log plot
displayed in Fig.~\ref{fig:Scaling} (left) this means that each
individual curve corresponding to a given value of $x=K-K_{c}$ can
be translated horizontally by some quantity $\ln\xi(x)$ {[}as $\ln(\xi t^{-1/3})=\ln(t^{-1/3})+\ln\xi${]},
so that all curves lie on a \emph{continuous} (except at $x=0$) $f(x)$
function. That this can indeed be done is illustrated in the center
plot of Fig.~\ref{fig:Scaling}. From this construction, one deduces
the values of $\xi(K)$, as shown in the right plot of Fig.~\ref{fig:Scaling}.
If there were an infinite number of noiseless curves, the only way
to match the horizontal critical curve $\Lambda(0)$ with the others
would be to displace it to infinity, showing, as one might have expected,
that $\xi(K_{c})$ shall diverge at $x=0$ as $(K-K_{c})^{-\nu}$,
and from this divergence the value of $\nu$ can be obtained. In the
imperfect, but real, experimental world, there is no such divergence,
so in practice we introduce a cutoff in the fitting function: $\xi(K)=\left[\alpha+\beta(K-K_{c})\right]^{-\nu}$.
The above method was developed to allow the determination of critical
exponents without achieving the thermodynamic limit. In the context
of the Anderson model, this limit means achieving very large number
of sites, and the method was dubbed ``finite-size scaling''~\citep{MacKinnon:OneParameterScalingOfConductance:PRL81,Pichard:FiniteSizeScalingAnderson:JPC:SSP81}.
In the present context, the ``thermodynamic limit'' corresponds
to very large \emph{times}, and the method, developed by G. Lemari\'e
and D. Delande, was called finite-\emph{time} scaling~\citep{Lemarie:These:09,Lemarie:AndersonLong:PRA09,Lopez:ExperimentalTestOfUniversality:PRL12}.

Using this procedure, the best value we obtained for the critical
exponent is $\nu_{\mathrm{exp}}=1.63\pm0.05$~\citep{Lopez:ExperimentalTestOfUniversality:PRL12}
that compares very well with the numerical value for the quasiperiodic
kicked rotor $\nu_{\mathrm{num}}=1.59\pm0.01$~\citep{Lemarie:UnivAnderson:EPL09}
and to the numerical value for the Anderson model $\nu_{\mathrm{num}}=1.571\pm0.008$~\citep{Slevin:AndersonCriticalExp:NJP2014}.

\subsection{Universality of the critical exponent}

The importance of critical exponents relies on the fact that they
are ``universal'', that is, they depend only on the symmetries of
the system, and not on microscopic details as the atom species, radiation
wavelengths, the path used for the measurement of the transition,
the values parameters as $\kbar$, $\omega_{2}$, $\omega_{3}$, etc.
The above value of $\nu$, around 1.6, is characteristic of the so-called
``orthogonal universality class'' of time-reversal-invariant systems~\citep{Haake:QuantumSigChaos:01}.
By varying some of the above ``microscopic'' parameters, and measuring
the critical exponent, we could perform an experimental test of this
universality~\citep{Lopez:ExperimentalTestOfUniversality:PRL12}.
Nine sets of parameters have been used, and the corresponding values
of $\nu$ range from 1.55 to 1.70, with an weighted average of 1.63$\pm$0.05.

\subsection{Study of the critical state}

\begin{figure}
\begin{centering}
\includegraphics[width=6.3cm]{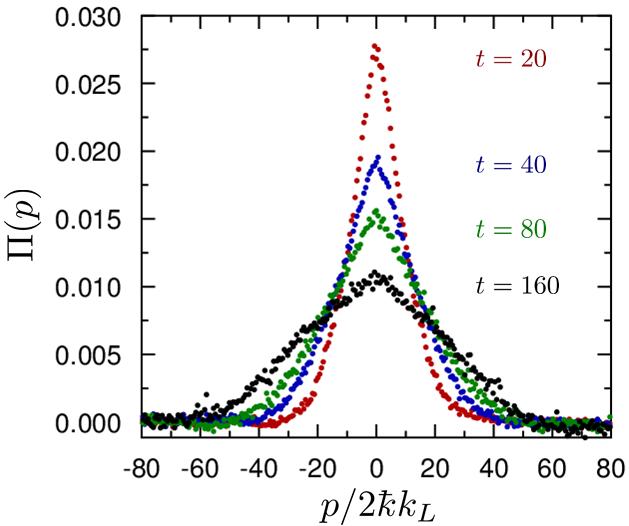}$\quad$\includegraphics[width=6.3cm]{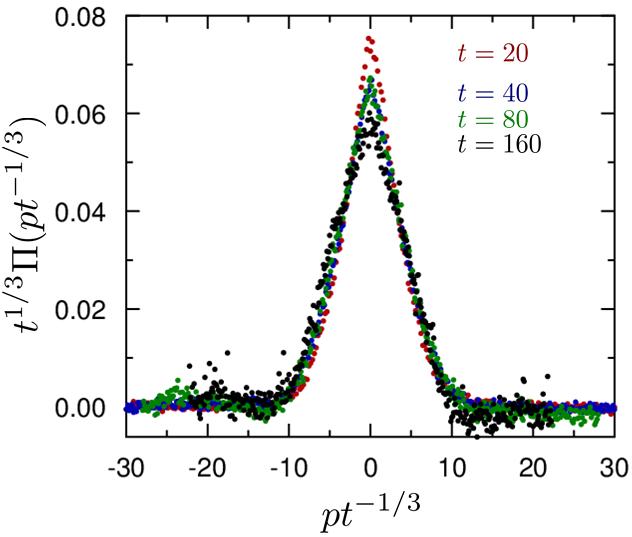}
\par\end{centering}
\caption{\label{fig:CriticalWaveFct}Experimentally measured critical wave
function (left), and its scaling (right). Residual decoherence effects
are responsible for the deviations from the perfect scaling, visible
on the top of the scaled distributions.}

\end{figure}

The critical state of a quantum phase transition presents distinctive
features characteristic of the transition. In condensed-matter physics,
the critical wave function is seldom accessible experimentally, but
this is not the case with cold atoms: In our system the momentum distribution
is directly measured~\citep{Lemarie:CriticalStateAndersonTransition:PRL10},
and even the complete wave function (i.e. including the phase information)
can, in principle, be measured. In the case of the Anderson transition,
the critical state is intermediate between a localized state, at the
high-disorder side \textendash{} or small $K$ for the kicked rotor
\textendash{} and a diffusive state at the low-disorder or high-$K$
side. One thus expects a subdiffusive behavior $\left\langle p^{2}\right\rangle \sim t^{\alpha}$
with $0<\alpha<1$; the scaling argument presented in sec.~\ref{subsec:AndersonTransition}
gives $\alpha=2/3$ (or, more generally, $\alpha=2/d$ in dimension
$d$). This behavior can also be seen in the momentum distributions:
If one performs the scaling $\Pi(p,t)\rightarrow t^{1/3}\Pi\left(pt^{-1/3}\right)$
the critical momentum distribution is expected to be invariant with
respect to $t$. This can be seen in Fig.~\ref{fig:CriticalWaveFct}.
It is worth noting that the distribution shape can be analytically
calculated from the self-consistent theory of Anderson localization,
and turns out to be an Airy function; this was also verified by comparison
with experimental distributions~\citep{Lemarie:CriticalStateAndersonTransition:PRL10}.

\subsection{Two-dimensional Anderson localization}

Dimension two is the ``lower critical dimension'' for the Anderson
model (cf.~sec.~\ref{sec:Anderson-model}). The eigenstates are
always localized, whatever the disorder strength, but the localization
length varies exponentially as $\ell\exp\left(\alpha k\ell\right)$,
where $k$ is the wave number of the propagating wave, $\ell$ the
mean-free path, and $\alpha$ a numerical constant of order of one.
For a 2D-Anderson-equivalent quasiperiodic kicked rotor, one has
\begin{equation}
p_{\mathrm{loc}}^{(2D)}=p_{\mathrm{loc}}\exp\left(\alpha\varepsilon(K/\kbar)^{2}\right)\label{eq:ploc2D}
\end{equation}
where $p_{\mathrm{loc}}=K^{2}/4\kbar$ is the 1D localization length.
This relation can be obtained by scaling arguments analogous to those
used in the context of Anderson localization. Moreover, the constant
$\alpha=\pi/\sqrt{32}$ can be determined analytically from self-consistent
theory, within (restrictive) assumptions. This makes the 2D behavior
very difficult to probe experimentally, as it implies observing large
localization lengths, and thus large localization times \textendash{}
which are limited in particular by the unavoidable presence of decoherence.
Recently, important experimental developments of our setup allowed
us to observe and study the 2D localization~\citep{Manai:Anderson2DKR:PRL15}.

\begin{figure}
\begin{centering}
\includegraphics[width=6.3cm]{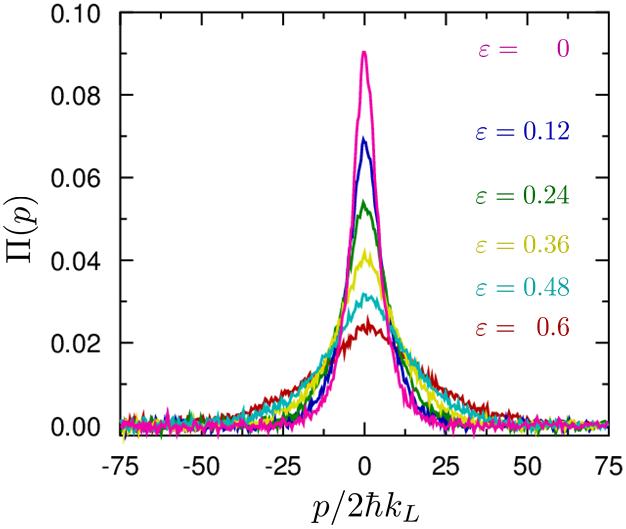}$\quad$\includegraphics[width=6.3cm]{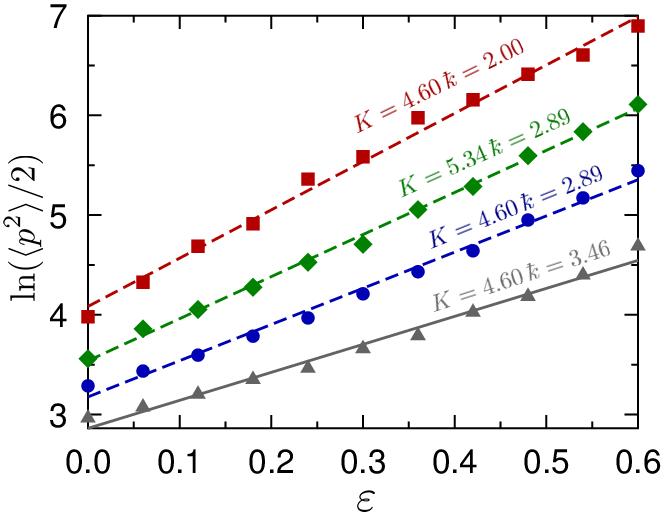}
\par\end{centering}
\caption{\label{fig:2D}Anderson localization in 2D. Left: Momentum distributions
at 1000 kicks, showing the crossover from the 1D ($\varepsilon=0$)
to the 2D behavior, characterized by a strong increase in the localization
length ($K=5.34$, $\kbar=2.89$). Right: Dependence of the log of
the saturated kinetic energy at $t=1000$ on the anisotropy parameter
$\varepsilon$, evidencing an exponential dependence as predicted
by Eq.~(\ref{eq:ploc2D}). One sees that the slope increases with
$K$ and decreases with $\kbar$; there is however a residual dependence
on $K/\kbar$ showing that $\alpha$ is slightly dependent on the
parameters (for more details, see ref.~\citep{Manai:Anderson2DKR:PRL15}).}

\end{figure}

In order to do so, we implemented experimentally the Hamiltonian
\[
H_{2D}=\frac{p^{2}}{2\mu}+K\cos x\left[1+\varepsilon\cos(\omega_{2}t+\varphi_{2})\right]\sum_{n}\delta(t-n)
\]
which maps onto a 2D Anderson model (provided $\omega_{2}/2\pi$ is
irrational, cf.~sec.~\ref{subsec:Equivalence3D}). By sweeping the
modulation amplitude $\varepsilon$ one can observe the crossover
from the 1D to the 2D behavior, Fig.~\ref{fig:2D} (left), which
manifests itself by an exponential increase in the localization length
according to Eq.~(\ref{eq:ploc2D}), as shown in Fig.~\ref{fig:2D}
(right).

\subsection{Conclusion}

The above results show a rather complete study of the Anderson model,
and in particular of the Anderson transition. This puts into evidence
the power of cold atom quantum simulators in general, and of the kicked
rotor in particular, to allow studies of phenomena extremely difficult
to observe with such a degree of precision and control in other contexts.

\section{Perspectives and conclusions}

In this paper we introduced basic ideas underlying the variants of
the kicked rotor used for the ``quantum simulation'' of the physics
of disordered systems. These ideas were illustrated with a series
of experimental results obtained along ten years of studies using
a cold-atom realization of this paradigmatic system. The wealth and
flexibility of the atomic kicked rotor are however not exhausted,
and other aspects of disordered quantum systems shall be explored
in the next years. Here are some possibilities:

\emph{i}) Localization is an interference effect. The system can evolve
along various paths going from an initial state $\left|\psi(t_{i})\right\rangle $
to a final state $\left|\psi(t_{f})\right\rangle $; if such an evolution
is coherent (i.e. if decoherence sources are well controlled) amplitudes
corresponding to different paths interfere. In the regime of weak
localization (short times, weak disorder), only a limited number of
simple paths contribute significantly. In a time-reversal-invariant
system one can show that loops leading back to initial state but described
in opposite senses have exactly the same phase, hence the corresponding
amplitudes interfere constructively. This effect has different manifestations:
One is the enhanced return to the origin, that is, $\left|\left\langle \psi(t_{f})\right.\left|\psi(t_{i})\right\rangle \right|^{2}$
is twice as large with respect to a system not time-reversal invariant.
Another is that a wave entering the disordered media along a direction
$\boldsymbol{k}_{0}$ is reflected along the reverse direction $-\boldsymbol{k}_{0}$
with an intensity twice as large as compared to other directions;
an effect called \emph{coherent backscattering}, which has been observed
with light~\citep{Wolf:WeakLocCBSLight:PRL85,Albada:ObsfWeakLocLight:PRL85,Labeyrie:CBSLightColdAtoms:PRL99}
and matter waves~\citep{Jendrzejewski:CBSUltracoldAtoms:PRL12,Cherroret:CoherentBackscatUltracoldMatter:PRA12}.
These effects can in principle also be observed with the kicked rotor.\emph{
Note added: }After this manuscript was submitted, our group observed
enhanced return to origin in the kicked rotor~\citep{Hainaut:ERO-KR:arXiv16}.

\emph{ii}) As evidenced by the preceding discussion, \emph{symmetries}
play a capital role in the physics of disordered/chaotic systems.
Spinless time-reversal-invariant systems belong to the so-called ``orthogonal
universality class'' and typically display 1D-localization, enhanced
return to the origin, and a phase transition in 3D with a critical
exponent around 1.6. This is the critical exponent we measured (sec.~\ref{subsec:AndersonTransition}),
definitely evidencing the time reversibility of our system (see also~\citep{Lignier:Reversibility:PRL05}
on the time reversibility of dynamic localization). Other universality
classes exist: The ``unitary'' class groups spinless systems that
are \emph{not} time-reversal invariant. As one can deduce from the
above discussion, they do not display coherent backscattering, and
the 3D critical exponent turns is around 1.44~\citep{Slevin:EstimateCriticalExpUnitary:arXiv16}.
One can break time-reversal invariance in the kicked rotor simply
by using a periodic kick sequence that has no (temporal) symmetry
axis. Its properties can thus be studied experimentally. The ``symplectic''
symmetry class is that of time-reversal-invariant systems \emph{with}
spin. It displays \emph{depleted,} instead of enhanced, return to
the origin and present a phase transition \emph{in 2D} with a critical
exponent 2.75~\citep{Slevin:AndersonCriticalExp:NJP2014}. Observing
such effects with the kicked rotor implies introducing some kind of
spin-orbit coupling. While this is conceivable~\citep{Scharf:KRForASpin1/2:JPAMG89}
it can be very difficult in practice. However, if this could be done,
it will open a wealth of new possibilities, as recent theoretical
suggestions including complex spin-orbit-coupled Hamiltonians allow
the realization of puzzling systems displaying \emph{momentum-space}
topological insulator properties and opening ways to the realization
of a Quantum Hall physics quantum simulator~\citep{Dahlhaus:QuantumHallEffectIQKR:PRB11,vanNieuwenburg:MetalTopologicalInsulatorQKR:PRB12,Chen:QuantumDrivenIntegerQuantumHallEffect:PRL14,Tian:EmergencerQuantumHallEffectChaos:PRB16}.

\emph{iii}) An intriguing phenomenon related to the Anderson physics
is \emph{multifractality}, which manifest itself especially at the
critical point of the Anderson transition. In such case, the critical
wave function (or the inverse participation ratio) present a\emph{
spectrum} of fractal dimensions with a characteristic log-normal distribution.
This phenomenon was observed with ultrasound waves~\citep{Faez:MultifractAnderson:PRL09},
but not with matter waves and can potentially be observed with our
system.

\emph{iv}) As it can be easily deduced from the discussion in sec.~\ref{subsec:Equivalence3D},
one can synthesize a Hamiltonian equivalent to a $d$-dimensional
Anderson model by using $d-1$ frequencies kick amplitude modulation.
This is easy to do experimentally, and opens the way to \emph{experimental}
studies of \emph{higher-than-three} dimensional Anderson models, and
to the determination of the so-called \emph{upper critical dimension},
for which the critical exponent of the transition coincides with the
prediction of the mean-field theory ($\nu=1$)~\footnote{The upper critical dimension is believed to be infinite.}.
Refs.~\citep{Ueoka:DimensionalDepfCriticalExpAndersonTr:JPSJ14,Slevin:EstimateCriticalExpUnitary:arXiv16}
give numerically calculated critical exponents for $d=4$ and 5. In
practice, however, measurement of the critical exponent for $d>3$
is not easy, because the characteristic times become very long. This
can be seen from Eq.~(\ref{eq:criticalp2}): Yet for $d=4$, the
typical evolution at criticality is $\left\langle p^{2}\right\rangle \sim t^{1/2}$.
We plan to measure the $d=4$ critical exponent in the near future
using a Bose-Einstein condensate, instead of simply laser-cooled atoms.

\emph{v}) The use of a Bose-Einstein condensate of potassium, that
is currently under development in our group, shall also allow us to
explore the effect of atom-atom interactions in a controlled way,
thanks to the so-called Feshbach resonances~\citep{Chin:FeshbachResonances:RMP10}.
The physics of disordered systems in presence of interactions \textendash{}
the so-called many-body localization \textendash{} is still largely
to be investigated~\citep{Schreiber:ObservationOfManyBodyLoc:S15}.
Numerical predictions for both the 1D kicked rotor~\citep{Shepelyansky:KRNonlinear:PRL93,Gligoric:InteractionsDynLocQKR:EPL11}
and for the Anderson model itself~\citep{Pikovsky:DestructionAndersonLocNonlin:PRL08,Flach:DisorderNonlin:PRL09}
indicate the existence of a sub-diffusive regime at very long times,
which has been observed in an experiment~\citep{Lucioni:SubdiffusionInteract:PRL11}.
A difficulty is that in the quasiperiodic kicked rotor localization
takes place in the momentum space, whereas contact interaction in
\emph{real} space translate into non-local interactions in \emph{momentum}
space. On the one hand, this makes the physics exciting, on the other
hand, this implies that the quasiperiodic kicked rotor with interactions
does not translate easily into a generalized ``Anderson'' model
with (local) interactions. Most numerical studies of the problem use
the approximation consisting in simply neglecting non-local effects
and keeping only the ``diagonal'' (local in momentum space) contribution~\citep{Shepelyansky:KRNonlinear:PRL93,Gligoric:InteractionsDynLocQKR:EPL11},
under the assumption that for weak enough interactions non locality
has negligible effects. To the best of my knowledge, there is no formal
justification, or even a careful numerical study of the validity of
this approximation. A numerically verified theoretical prediction
(using this approximation) for the quasiperiodic kicked rotor indicates
that the metal-insulator transition survives, but the localized regime
is replaced by a sub-diffusive one~\citep{Cherroret:AndersonNonlinearInteractions:PRL14}.

The study of Anderson physics is far from exhausting the possibilities
of the kicked rotor as a quantum simulator. The kicked rotor and related
dynamical systems can also be mapped onto other condensed-matter systems,
e.g. the Harper model and its famous ``Hofstadter butterfly''~\citep{Gong:HarperKR:PRA08}.

In conclusion, the best is still to come!

\section{Acknowledgments}

The work described in this article is a team work. If I had the honor
to receive the ``Leconte prize'' of the Acad\'emie des Sciences,
it is the work of the team that deserved it. I am very happy to have
this opportunity to acknowledge it. I most warmly thank the people
with whom I had the pleasure of collaborating in the last 20 years,
both at the PhLAM laboratory in Lille and at the Kastler-Brossel laboratory
in Paris. I am particularly grateful to the numerous PhD students
\textendash{} who are the true motor of the research \textendash{}
with whom I could work. Various funding agencies also contributed
to make it possible, in particular the Centre National de la Recherche
Scientifique, Agence Nationale de la Recherche (Grants MICPAF No.
ANR-07-BLAN-0137, LAKRIDI No. ANR-11-BS04-0003 and K-BEC No. ANR-13-BS04-0001-01),
the Labex CEMPI (Grant No. ANR-11-LABX-0007-01), and ``Fonds Europ\'een
de D\'eveloppement Economique R\'egional'' through the ``Programme
Investissements d'Avenir''.

\bibliographystyle{elsarticle-num}

\end{document}